\documentclass[letterpaper]{article} 
\usepackage{aaai24}  
\usepackage{times}  
\usepackage{helvet}  
\usepackage{courier}  
\usepackage[hyphens]{url}  
\usepackage{graphicx} 
\usepackage{natbib}  
\usepackage{caption} 
\usepackage{algorithm}
\usepackage{algorithmic}

\usepackage{newfloat}
\usepackage{listings}
\DeclareCaptionStyle{ruled}{labelfont=normalfont,labelsep=colon,strut=off} 
\floatstyle{ruled}
\newfloat{listing}{tb}{lst}{}
\floatname{listing}{Listing}
\title{Introducing ELLIPS: An Ethics-Centered Approach to Research on LLM-Based Inference of Psychiatric Conditions}
\author{
    Roberta Rocca\textsuperscript{\rm 1}\equalcontrib,
    Giada Pistilli\textsuperscript{\rm 2}\equalcontrib,
    Kritika Maheshwari\textsuperscript{\rm 3},
    Riccardo Fusaroli\textsuperscript{\rm 1}
}
\affiliations {
    \textsuperscript{\rm 1}Aarhus University \\
    \textsuperscript{\rm 2}Hugging Face \\
    \textsuperscript{\rm 3}Delft University of Technology\\
    roberta.rocca@cas.au.dk, giada@huggingface.co, k.maheshwari@tudelft.nl, fusaroli@cc.au.dk 
}

\usepackage{bibentry}

\begin{document}
\maketitle
\begin{abstract}
As mental health care systems worldwide struggle to meet demand, there is increasing focus on using language models (LM) to infer neuropsychiatric conditions or psychopathological traits from language production. Yet, so far, this research has only delivered solutions with limited clinical applicability, due to insufficient consideration of ethical questions crucial to ensuring the synergy between possible applications and model design.
To accelerate progress towards clinically applicable models, our paper charts the ethical landscape of research on language-based inference of psychopathology and provides a practical tool for researchers to navigate it. We identify seven core ethical principles that should guide model development and deployment in this domain, translate them into ELLIPS, an ethical toolkit operationalizing these principles into questions that can guide researchers' choices with respect to data selection, architectures, evaluation, and model deployment, and provide a case study exemplifying its use. With this, we aim to facilitate the emergence of model technology with concrete potential for real-world applicability.
\end{abstract}

\section{Introduction}
\subsection{Inferring Psychopathology From Language}
Over the past decades, a growing body of research has focused on identifying digitally trackable markers of neuropsychiatric conditions that can be used to support detection, assessment, and monitoring, including in remote settings \cite{insel_digital_2018, vasudevan_digital_2022}. Identifying reliable digital markers would carry great significance for mental health care systems worldwide. Mental health care systems are struggling to meet fast-growing demand, with an estimated 1 in 8 people worldwide suffering from a psychiatric condition, and only 10\% of them being able to access mental health care services, which has severe social and economic implications \cite{world_health_organization_world_nodate, united_nations_development_programme_human_2022, plana-ripoll_comprehensive_2019,saxena_excess_2018,thornicroft_premature_2013,plana-ripoll_exploring_2019}.

Language is receiving increasing attention as a possible source of such digital markers \cite{corona_hernandez_natural_2023}. Language production encodes rich information about individual traits (e.g., personality, \citealp{yarkoni_personality_2010,park_automatic_2015}), mental states and psychopathology \cite{nguyen_affective_2014,williamson_detecting_2016}, both in its content (what we talk about) and its style (e.g., lexical, syntactic, and discourse-level choices in \textit{how} we talk about it) \cite{rocca_language_2022}. Indeed, language is central in psychiatric assessment and diagnostics, and diagnostic criteria for many conditions include symptoms that are primarily inferred from linguistic behavior \cite{low_automated_2020}. For example, a primary diagnostic criterion for schizophrenia is the presence of thought disorder symptoms such as poverty of speech or tangentiality, which are primarily inferred from patients' speech \citep{kircher2018formal}.

\subsection{The Promise of (Large) Language Models}
Research on linguistic markers of neuropsychiatric conditions has traditionally focused on identifying statistical associations between quantitative descriptors of semantic and stylistic aspects of language production in linguistics tasks (e.g., measures of semantic coherence in autobiographical narratives) and psychiatric diagnoses (whether that person has received a diagnosis for a specific condition, e.g., schizophrenia, see \citealp{parola2023speech}). More recently, the field has started shifting its focus from statistical modeling (aimed at identifying linguistic markers) to predictive modeling (aimed at inferring conditions), and from the use of theory-driven input measures to complex input representations such as vectors \cite{ciampelli2023combining}, \cite{hansen_speech-_2023}. This prediction-oriented shift aims at delivering models with applications in supporting and augmenting clinical diagnostics and assessment processes.\cite{yarkoni_choosing_2017,shmueli_explain_2010,rocca_putting_2021}.

The advent of transformers, \cite{vaswani_attention_2017,devlin_bert_2019} and more recently, of large language models (LLMs) has dramatically increased the potential of this line of research. These models can learn complex relations between language and psychiatric conditions, and \textit{transfer learning} \citep{ruder_transfer_2019} makes possible to develop high-performing model even with little data, which is vital considering the scarce public availability of clinical data.

\subsection{Mismatch With Clinical Goals}
While language models open up a vast space of possibilities, research on language-based inference of neuropsychiatric conditions has hitherto failed to deliver systems with concrete practical applicability.

Arguably, this is largely due to predictive modeling problems not being more deeply informed by clinical knowledge and practices. For example, most research on linguistic markers of neuropsychiatric conditions focuses on developing diagnosis-specific binary classification models that discriminate between people who have a clear diagnosis without co-morbidities or other complications, and people without psychiatric conditions, nor symptoms thereof. Besides providing initial proof of concept for potential markers, this approach quickly becomes unproductive. Diagnoses are deeply useful pragmatic tools, but their biological nature is increasingly questioned. Indeed, there is extreme heterogeneity in symptom profiles \textit{within} diagnostic groups, substantial overlap in symptom profiles \textit{across} diagnoses, and no distinct biological underpinnings for different diagnoses \cite{roefs_new_2022, conway_hierarchical_2019, fried_depression_2015,cross-disorder_group_of_the_psychiatric_genomics_consortium_electronic_address_plee0mghharvardedu_genomic_2019,galatzer-levy_636120_2013}.  
As shown in a recent study \cite{hansen_speech-_2023}, this is reflected in huge performance differences between models trained to discriminate between individuals with a given diagnosis and ``clean'' controls (with reported F1-scores of up to .9 for some diagnoses), and models trained to discriminate between \textit{multiple} diagnoses with partly overlapping symptomatology, where performance levels are much lower (i.e., in the study at stake, no higher than .5). This is also particularly relevant for real-world clinical decision-making. In fact, in clinical settings, clinicians are asked to discriminate between a wide range of possible diagnoses, rather than making binary decisions on the presence or absence of an individual diagnosis \cite{corona_hernandez_natural_2023, elvevaag2023reflections}.

This situation does not solely arise from insufficient engagement with knowledge from psychiatric nosology and with the current transdiagnostic turn (which advocates for a focus on inference of clinical traits relevant for multiple diagnoses). The lack of experimentation with more ``realistic'' predictive modeling setups also arises from data scarcity. Partly due to the sensitive nature of linguistic and psychopathological data, there are currently no public large-scale datasets including both records of text or speech production and clinical phenotypes from multiple diagnostic groups.

\subsection{An Ethical Compass for Impactful Research}
These considerations highlight how opportunity (e.g., which data and target variables are easy to access), rather than the maximization of scientific and/or societal utility of the resulting systems, has so far informed the choice of predictive modeling targets and the development of underlying systems. Failing to engage with knowledge of psychiatric nosology and with considerations related to clinical decision-making is diverting resources from the applied focus of the field.

Insofar as they directly affect models' potential for real-world applicability and scientific insight, ``technical'' choices concerning model development and deployment (e.g., the selection of the target variables, as well as choices concerning the use of data, model architecture, training, and evaluation procedures) have an intrinsic \textit{ethical} dimension. For the field to produce research that can deliver on its ambition to yield an overall positive impact, a radical change in research practices is needed which places ethical considerations at the center of decisions concerning all stages of model development and deployment.

Crucially, the ethical landscape of research on language-based inference of psychopathology is extremely complex, as this research involves a compound ecosystem of stakeholders which includes vulnerable individuals, knowledge of elaborate decision-making and responsibility structures involved in clinical practice, and an underlying behavioral signal which is widely available online, which increases the potential for dual use of resulting models.
Expecting ML practitioners to navigate it in the absence of guidelines that help them orient themselves through its complexities is unreasonable and unrealistic, and, at present, work devoted to supporting them is sparse and unsystematic. The present paper aims to provide fundamental contributions towards filling these gaps.

\section{Aims and Scope}
With this paper, we aim to provide a first systematic exploration of the ethical dimensions involved in the development and deployment of systems for language-based inference of neuropsychiatric conditions and related psychopathology. We define research on language-based inference of (neuro)psychiatric conditions as all research concerned with developing systems that can infer psychiatrically relevant individual traits from linguistic input, be it text or speech. Importantly, we focus on identifying ethical considerations that should inform decisions concerning \textit{technical} aspects of the model development and deployment pipeline conditional on a clear specification of the intended potential applications of the resulting model. This is a crucial step, as model design is tightly related to the intended goals of the model. We articulate these into two main stages: model \textbf{development} and \textbf{post-development} phases, which includes deployment.

With respect to development, ML practitioners make decisions on four overarching (and intrinsically related) questions:
\begin{itemize}
\item {\textbf{Target variables}}: deciding which kind of trait the system tries to predict based on representations of linguistic behavior (e.g., diagnoses, transdiagnostic cognitive and psychopathological traits, fine-grained measures of symptom severity);
\item {\textbf{Training data}}: deciding what kind of linguistic behavior is fed as training data to the model (e.g., text or speech, naturalistic or task-based production), which population the data is drawn from (e.g., clinical or general population, single \textit{versus} multiple diagnostic groups, sample diversity), and potentially collecting this data;
\item {\textbf{Model architecture}}: deciding which kind of model is used to perform inference (type of architecture and size), if its weights are or should be pretrained, whether predictions are produced in a generative fashion or with traditional inference modules (e.g., classification layers in a neural network), and how to train it;
\item {\textbf{Evaluation protocols}}, deciding which data is used to evaluate the system; which metrics are used to evaluate the system; and whether and how to explain model predictions.
\end{itemize}

With respect to post-development phases, researchers are faced with decisions concerning two main classes of problems:
\begin{itemize}
\item {\textbf{Model and data sharing}}: deciding whether training data and model weights and architecture are made available and in what form (e.g., open-source, upon request, gated access);
\item {\textbf{Mode of deployment}}: identifying how the resulting system could be integrated into clinical practice (e.g., how model predictions interact with human expertise, how the model is served to ensure that clinicians understand it and are aware of limitations).
\end{itemize}

Our goal is two-fold. First, we aim to identify and describe ethical principles (informed by previous work in research ethics, biomedical ethics, and machine learning ethics) that should inform decisions across these six stages, and explain their relevance with respect to individual stages. Secondly, we aim to translate these principles into a set of concrete questions that researchers can use to inform and scaffold the development of their predictive modeling systems. These result in \textbf{ELLIPS} (an \textbf{E}thics-centered approach to research on \textbf{LL}M-based \textbf{I}nference of \textbf{PS}ychiatric conditions), a toolkit that can concretely help researchers place ethical considerations at the center of their ML development work.
To demonstrate the utility of the resulting toolkit, we will show it in action by presenting a case study targeting research on language-based inference of autism, a high-incidence neuropsychiatric condition. With this, we hope to encourage the adoption of ethics-centered research practices that will deliver systems with concrete potential for positive impact not only within this field, but more generally across subfields of machine learning and natural language processing concerned with applications of language technology for positive societal impact.

\section{Ethical Principles}
Ethical principles generally relevant for research on language-based inference of psychiatric conditions and their psychopathology can be identified based on previous work in at least three relevant domains. \textit{Scientific research ethics }is a subdomain of ethics governing the standards of conduct and integrity for scientific researchers. \textit{Biomedical ethics} is concerned with ethical questions arising in biomedicine and medical research and practice. \textit{Machine learning ethics} (or ML ethics) is a rapidly developing domain concerned with ethical questions arising in the development and deployment of machine learning systems, and their impact on humans. 

These three domains are deeply interwoven. While scientific research ethics identifies domain-general principles of ethical practices in research, biomedical ethics, and machine learning ethics interpret these in ways that are relevant for specific domains. In our attempt to describe the ethical landscape of research on language-based inference of neuropsychiatric conditions, we refer to pivotal work in biomedical ethics and ML ethics (as well as to domain-specific knowledge) to operationalize core principles of scientific research ethics in ways that are relevant for our target domain.

For scientific research ethics, we will rely on principles identified by the \textit{Belmont Report}\footnote{Available at: \url{https://www.hhs.gov/ohrp/regulations-and-policy/belmont-report/read-the-belmont-report/index.html}} and by \citet{douglas2014moral}. First drafted in 1978, the Belmont Report was created by the National Commission for the Protection of Human Subjects of Biomedical and Behavioral Research to summarize ethical principles and guidelines for human subject research, placing special emphasis on the direct and indirect impact of research on human subjects. The Belmont Report outlines three unifying ethical principles for human subjects research: \textbf{respect for persons} or \textbf{autonomy}, \textbf{beneficence}, and \textbf{justice}, whose relevance for research on language-based inference of psychiatric conditions will be discussed in the next section. For biomedical ethics, we will refer to Tom Beauchamp and James Childress's seminal work Principles of Biomedical Ethics \cite{beauchamp1978principles}, which identifies four principles: autonomy, justice, beneficence, and non-maleficence. The first three principles fully overlap with the principles outlined in the Belmont Report, while the final principle draws its roots from Hippocrates' ``Do no harm''. As we will specify in the dedicated subsection, in line with the Belmont report, we combine beneficence and non-maleficence under the same principle.

In addition to the three principles from the Belmont report, we discuss four principles introduced by \citet{douglas2014moral}'s work on moral dimensions of scientific research, namely \textbf{responsible scientific inference}, \textbf{credit allocation}, \textbf{transparency}, and \textbf{social responsibility}. In the next subsections, we will introduce each of these seven principles and discuss their relevance to our target domain. To highlight the relevance of each principle to specific stages of model development, we mention relevant stages in bold and between squared brackets.

\subsection{Autonomy}
The principle of \textbf{respect for persons} or \textbf{autonomy} advocates treating individuals as autonomous agents capable of making decisions for themselves, and making sure that those decisions are respected by others. This implies a duty to provide information necessary for informed decision-making, and the need to protect those with diminished autonomy, such as individuals with mental disabilities, illnesses, or those in restrictive circumstances. In the context of scientific research, this principle aims to ensure that participants enter an experiment voluntarily and provide their \textit{informed consent} armed with adequate information on the scope and the implications of the research.

In the context of biomedical ethics, autonomy involves respecting the patient's autonomous capacity to make informed decisions on their health \textit{without controlling influences}. It is particularly important here that patients are not forced by external pressures -- e.g., from physicians, family members, partners, or even technologies -- to make decisions about their health (e.g., testing, treatments) that they would not have otherwise made. Confidentiality and informed consent are also relevant to this principle. Patients must receive timely and truthful information about healthcare procedures (and who will access and use their health information and how) in a way that is accessible to them.

For research on language-based prediction of neuropsychiatric conditions, the principle of autonomy is relevant across multiple stages of design and deployment of a predictive modeling systems. With respect to the selection of {\textbf{training data}}, regardless of whether new data collection is involved or previously collected data are re-used, researchers need to ensure that participants (including potentially vulnerable ones) provide (or have provided) their consent to participation robustly, clearly, and comprehensively.

Furthermore, participants need to be adequately informed of the scope and implications of the research and of how their data will be used and by whom, and whether and how (e.g., anonymously) they will be made publicly available [\textbf{{model and data availability}}; \textbf{{mode of deployment}}]. Note that an important challenge in this respect concerns understanding whether training data can be retrieved or reconstructed from model weights \cite{nasr2023scalable}.

Considerations related to consent are also crucial with respect to clinical deployment. Patients for whom clinical assessment involves the use of a predictive model should be informed about how the system formulates predictions [\textbf{{target variables}}, \textbf{{model architecture}}], what their performance limitations are [\textbf{{evaluation protocols}}], how they are used for decision-making [\textbf{{mode of deployment}}] in an intelligible and accessible way. 

Finally, the principle of autonomy could also be interpreted as advocating for active participation of patients in informing research goals and objectives, for example through the involvement of patients, their communities, and associations, along with clinicians, in research design.

\subsection{Beneficence}
The second principle of the Belmont report is the principle of \textbf{beneficence}, which underscores the moral obligation to protect individuals from harm, and actively secure their well-being. In the formulation from the Belmont report, this principle encompasses two fundamental rules: the duty to \textit{do no harm}, a fundamental principle in medical and research ethics, and the \textit{obligation to maximize possible benefits while minimizing potential harms}, which poses the challenge of balancing the pursuit of beneficial outcomes against possible risks. 

This principle combines those of \textbf{beneficence} and \textbf{non-maleficence} from biomedical ethics \cite{beauchamp1978principles}. In this context, these principles emphasize that medicine should primarily aim to benefit the patient while also striving to prevent or reduce harm, which often implies trade-offs between the two. For example, while surgery sometimes involves cutting the skin with a scalpel, this harm is offset by the good that the surgery would have on the patient's health and well-being. The principle of beneficence and non-maleficence can also also be interpreted in relation to the society as a whole from a public health perspective. For example, while vaccines involve a financial cost to society, as well as a sometimes painful needle jab to individuals, they decrease the spread of communicable illness and help protect immunocompromised members of society through herd immunity.

The principle of beneficence has deep implications for research on language-based inference of neuropsychiatric conditions. Here, harms associated with the development of these systems are related to errors in models' predictions (which may result in inadequate treatment), or to their potential for dual use (i.e., models being used to profile and target individuals). Benefits, on the other hand, concern increased precision and scalability in assessment and monitoring of psychiatric conditions, which can deliver more personalized interventions and increase the availability of resources for individualized support. At a broader level, benefits include the alignment of capacity with demand in mental health care systems.

A key goal of research on language-based inference of neuropsychiatric conditions is thus to maximize the potential to deliver these benefits (e.g., supporting assessment or remote detection and monitoring of conditions), while minimizing the potential harms (e.g., consequences of inaccurate predictions, such as wrong treatment recommendations). Doing so involves making adequate choices at all stages of model design. For example, models should be optimized so that the risk of incorrect predictions on real-world data is low, which requires, for example, using {\textbf{training data}} (and evaluation) of adequate size and level of heterogeneity, and as similar as possible to the kind of data that would be collected to deploy this system in a clinical setting. The choice of \textbf{{model architectures}} and training protocols of sufficient complexity and robustness to idiosyncratic variation in the data should be preferred. Finally, \textbf{{evaluation protocols}} should be able to adequately measure performance in ways that make it possible to assess the expected real-world performance of the model.

An important additional set of considerations concerns models' affordances for use in clinical settings. The choice of target variables should ensure that models can be beneficially deployed to \textit{complement} and \textit{inform} clinicians' judgment, rather than \textit{replace} it [\textbf{{target variables}}]. For example, predicting diagnoses (rather than, for example, inferring fine-grained symptom or clinical test scores) is arguably not going to be particularly useful in a clinical context. It is in fact unrealistic and ethically questionable to replace clinicians' judgment about diagnoses or relevant interventions, grounded on expert knowledge about patients' history, with model predictions \citep{grote2020ethics, tobia2021does}. Furthermore, to be beneficially deployed in clinical settings, models should be complemented with \textit{explainability} toolkits that make it possible to responsibly deploy them in clinical decision-making, satisfying the need for accountability intrinsic to clinical practice [\textbf{{mode of deployment}}]. 

Finally, to minimize the risk of harms related to dual use while still preserving the knowledge benefits of open sharing of scientific resources, researchers should carefully consider how models and data can be made available to decrease probability of undesired uses [\textbf{{model and data sharing}}]. 

\begin{figure*}[htpb!]
  \includegraphics[width=1.01\textwidth]{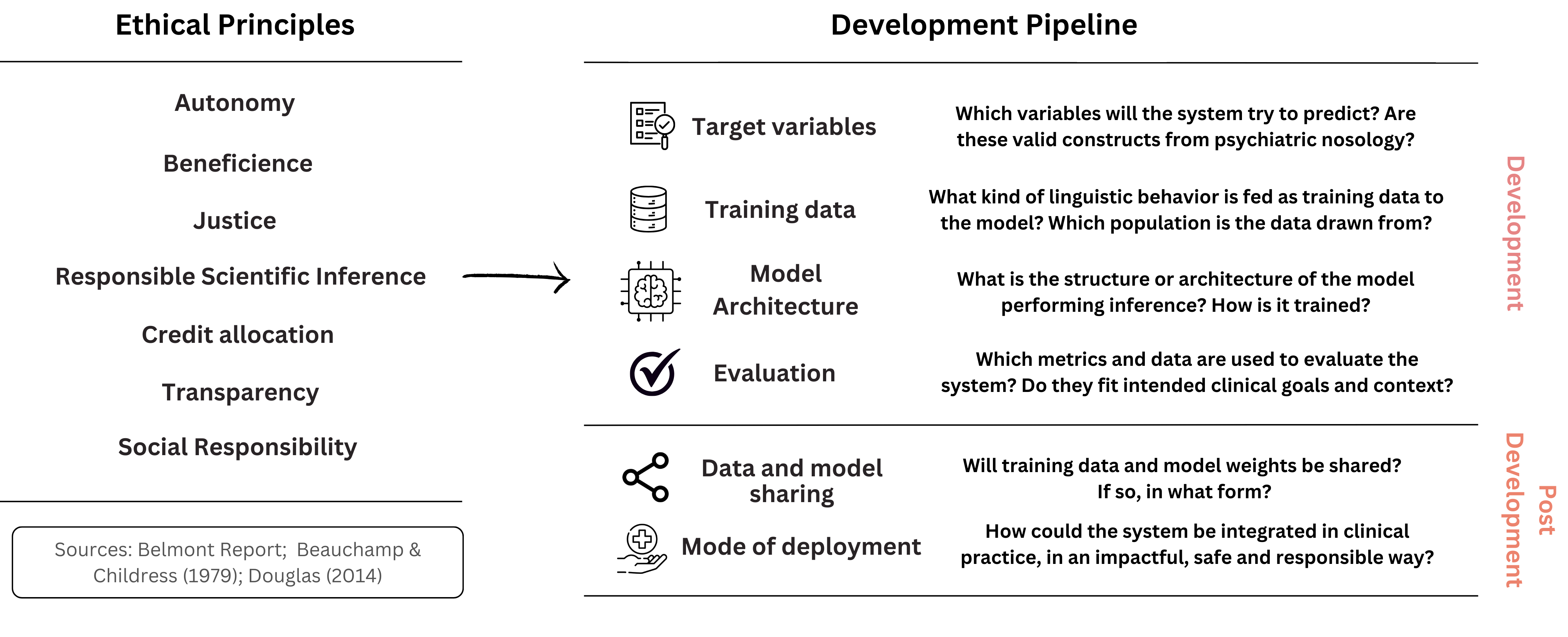}
  \caption{Visual overview of ethical principles informing stages of model development}
\label{figure1}
\end{figure*}

\subsection{Justice}
The third principle of the Belmont report, the principle of \textbf{justice}, emphasizes fairness in distributing the benefits and burdens of research. Historically, research has involved practices where more privileged segments of society accrued the benefits of research, while disadvantaged groups mostly carried the burdens. This is also true of LLM research, with the overwhelming majority of LLM technology being developed for high-resource languages \citep{joshi2020state}.

In biomedical ethics, the principle of justice \cite{beauchamp1978principles} is often viewed through the lens of distributive justice; that is, ensuring that there is a fair distribution of healthcare goods and services. Healthcare testing and procedures should do their best to ensure fair outcomes for all patients. Research and recommendations surrounding their use should take into consideration their efficacy within different patient demographics and characteristics.

In the context of research language-based inference of psychiatric conditions, this principle has deep implications for several stages of model development. Given that language is affected by dimensions such as diagnostic groups, gender, ethnicity, age, socio-economic background and multilingualism,  researchers should strive to develop systems that ensure fair predictive performance across these dimensions. To do so, choices related to data representation at training [{\textbf{training data}}, {\textbf{model architectures}] are crucial, along with {\textbf{evaluation protocols}} that allow for a fine-grained understanding of potential performance disparities across groups.

Data representation \cite{panch2019artificial, baird2020considerations} and the associated risk of biases \cite{parikh2019addressing, panch2019artificial} directly impact the model's fairness and decision-making abilities. Both underrepresentation and overrepresentation of certain groups in the data can lead to uneven outcomes. For instance, when a particular demographic is overrepresented in the dataset, there is a risk that the model becomes biased towards the characteristics of that group (for example, always predicting a given diagnosis for a specific demographic group). Conversely, underrepresentation can lead to certain groups being virtually invisible to the model, resulting in its failure to accurately or adequately address their specific needs or conditions. The issue of representation becomes even more complex when training data are sourced from the internet, which does not proportionately represent all segments of society or humanity. Certain groups may be more active or visible online and speak more one language than another, while others may have limited digital footprints due to various socioeconomic, cultural, or geographic factors. Therefore, datasets scraped from the internet can inherently carry these imbalances, which, if not addressed, lead to ML models that are skewed and potentially unfair.

Besides distributive concerns, the principle of justice also extends to concerns around epistemic injustice. It focuses on systematically excluding certain patient groups from contributing to testimonial or hermeneutical resources for psychiatric classification due to wrongly presumed irrelevance of their knowledge or expertise \citep{bueter2019epistemic}. Relatedly, the principle of justice also emphasizes the need to focus on developing systems that do not only serve privileged language groups (i.e., English-speaking, see \citealp{peled2018language}). This involves, for example, an obligation to strive to develop and share resources (e.g., datasets, {\textbf{model and data sharing}}) that can be used to build models that serve more and lesser-resourced languages -- which is also a precondition for drawing truly generalizable scientific inference on mechanistic relations between language and psychopathology. 

\subsection{Responsible Scientific Inference}
The principle of \textbf{responsible scientific inference}, which highlights the need to assess and carefully weigh the potential consequences of errors in our design choices. Although the principle is primarily developed with experimental research in mind, in our domain of interest this principle can be translated into the need to develop \textbf{{evaluation protocols}} and explainability toolkits that allow for detailed understanding and careful interpretation of model predictions (e.g., communicating confidence). Additionally, this principle can be interpreted as suggesting that researchers should strive to make it possible for others to further evaluate and stress-test the models, for example by sharing the model, data, and relevant code [\textbf{{model and data availability}}].

\subsection{Credit Allocation}
Secondly, \citet{douglas2014moral} highlights the imperative of \textbf{appropriate credit allocation}, which stresses the need to ensure that all scientific sources relevant to a given research endeavor are appropriately incorporated and credited. This principle is especially important for our field of research, as it implies the need to accurately incorporate knowledge from both machine learning and psychiatric nosology in the design of a system. Going back to an issue that has already partly been discussed, this involves, for example, considering whether the choice of \textbf{{target variables}} (e.g., categorical indicators of diagnoses) aligns with state-of-the-art knowledge on psychiatric nosology and clinical practice.

\subsection{Transparency}
Thirdly, \citet{douglas2014moral} highlights the principle of \textbf{transparency} \cite{bosma2022sharing, sim2006clinical}, which requires peer evaluation of evidence and replication of results. Transparency ensures that the scientific community and the wider public can trust and verify the work being conducted. Understanding and documenting the origins of the {\textbf{training data}}, or \textit{data provenance}, is critical to this principle. Researchers must ensure transparency \cite{larsson2020transparency} and traceability \cite{mora2021traceability} of data sources. This also poses an obligation to regard (wherever possible, e.g., with respect to considerations related to consent and confidentiality), {\textbf{data and model sharing}} as necessary. In addition to this, the principle of transparency dictates that researchers adequately report performance in ways that provide a comprehensive picture of models' limitations [\textbf{{evaluation protocols}}]. 

\subsection{Social Responsibility}
Finally, \citet{douglas2014moral} identifies \textbf{social responsibility} as a core principle of ethical research. This emphasizes scientists' moral duty to reflect on the social implications of their research \cite{douglas2014moral}, which involves assessing both the potential harms and benefits of their work, as well as effectively communicating their findings to the broader public in a manner that is both comprehensible and accessible. This principle is dense with implications for research on language-based prediction of psychiatric conditions and it is closely related to the Belmont report's principle of beneficence (similarly concerning all stages of model development and post-development) as it calls for researchers to strive to design systems that can positively contribute to mental health care individually and at scale, while minimizing the potential for dual use.

\section{ELLIPS: A Novel Ethical Toolkit}
In the previous section, we highlighted several core principles of research ethics, biomedical ethics, and machine learning ethics and their relevance across the \textbf{development} and \textbf{post-development} phases of the machine learning pipeline. We focused especially on four core problems in ML development, namely the selection of \textbf{{target variables}} (which should be informed by considerations related to the intended real-world applications of the model, as well as its adherence to knowledge in relevant fields), the identification or collection of \textbf{{training data}} (where issues like data provenance, consent, and representation emerge), the selection and training of \textbf{{model architectures}} (which brings into question the ethics of algorithmic choices, risks of reinforcing biases present in the training data, and questions related to interpretability), and the development of suitable \textbf{{evaluation protocols}} (which involves devising batteries of tests that allow to iteratively evaluate performance and understand strength and limitations in real-world deployment settings). 

We also observed how post-development choices such as decisions concerning \textbf{{model and data sharing}} are related to core ethical principles, such as transparency and the obligation to engage in responsible scientific inference. Finally, the principles reviewed above highlight the importance of questions concerning the \textbf{{mode of deployment}}, that is, imagining whether and how the resulting system could be embedded in clinical decision-making.

The resulting body of considerations is theoretically dense and highly complex, and a crucial question relates to how to distill it into a set of \textit{pragmatic} considerations, or a ``checklist'' that researchers can use to inform the development of an ethically robust systems for language-based inference of psychiatric conditions and their psychopathology. This is the question we address in this section by presenting \textbf{ELLIPS} (an \textbf{E}thics-centered approach to research on \textbf{LL}M-based \textbf{I}nference of \textbf{PS}ychiatric conditions), a toolkit consisting of a number of core ethical questions that ML practitioners and researchers can use to inform their decisions concerning all stages of the development of a predictive system.

\subsection{Target Variables}
With respect to the selection of target variables, the principles discussed in the previous section raise the following core questions:
\begin{itemize}
\item  Which target variables are we trying to predict? How are they \textit{useful} with respect to the intended application of the model? For example, are they relevant from a clinical perspective, i.e., can a system predicting these variables be used to complement, rather than replace or subtract responsibility from, human judgment? [\textbf{beneficence}, \textbf{social responsibility}]
\item What are the downsides of these variables, i.e., could a model predicting those be easily used for unintended, harmful uses (i.e., discriminatory ones)? [\textbf{beneficence}, \textbf{social responsibility}, \textbf{justice}]
\item Are we trying to predict target variables that are \textit{scientifically} robust and grounded in reliable and valid constructs (i.e., supported by psychiatric nosology and psychometrics research)? What are the implications if they do not? [\textbf{responsible scientific inference}, \textbf{credit allocation}]

\end{itemize}

\subsection{Training Data}
With respect to the selection or collection of training data, based on the principles highlighted above, we identify the following questions: 
\begin{itemize}
\item If existing data sets are used, can we verify and document the origins of this data (\textit{data provenance})? [\textbf{autonomy}; \textbf{transparency}] 
\item If existing resources are used, how do we approach the challenge of obtaining informed consent, and/or navigating potential conflicts between the original intent of the data and its current use? [\textbf{autonomy}]
\item If new data are collected, what methods or frameworks are we employing to guarantee informed consent? In cases where research subjects are vulnerable (due to factors like mental health conditions), what additional measures are taken to ensure their protection and ability to provide consent? [\textbf{autonomy}]
\item If new data are collected, is the data collection protocol sufficiently documented so to guarantee full transparency, both with respect to ethical questions, and with respect to the implications of data collection on model performance and potential biases? [\textbf{transparency}, \textbf{justice}]
\item Are community voices and perspectives incorporated into the research process to ensure relevance and fairness? [\textbf{autonomy}, \textbf{justice}, \textbf{social responsibility}]
\item If existing data is used, how do we ensure that the data accurately represents the diverse populations and groups (e.g., with respect to ethnicity, age, immigrant status, linguistic variations) affected by psychiatric conditions? If new data are collected, what criteria are used to determine the representation of diverse groups in the research, including but not limited to different racial, ethnic, and socio-economic backgrounds? [\textbf{justice}]
\item Is the data multilingual, so to guarantee that the benefits of this research do not \textit{only} extend to a single language group? If not, can data collection target a multilingual population? If monolingual, does the data represent sufficient variations in language use within that linguistic population (e.g., accents, dialects, multilingual speakers), so to prevent biased performance with respect to specific language subgroups? [\textbf{justice}]
\item How do we ensure that the selection of research participants is equitable and not biased towards easily available or vulnerable populations (e.g., students, underpaid labor), both in the interest of robustness and generalization, as well as in the interest of social justice? [\textbf{justice}]
\end{itemize}

\subsection{Model Architecture}
With respect to the design and selection of model architectures, which should balance technical capabilities and adherence to ethical imperatives, the following questions emerged from our review of ethical principles in the previous section:
\begin{itemize}
\item Why is the type and size of the chosen architecture expected to provide adequate, reliable, and generalizable predictive performance across diverse groups given the nature and size of the training data and of the data expected to be used at deployment? [\textbf{beneficence}, \textbf{social responsibility}, \textbf{responsible scientific inference}] 
\item Are the type and size of architecture compatible with the use of techniques that make it possible to \textit{explain} model decisions (e.g., provide confidence estimates, or ground decision in input components), and communicate them to all stakeholders (e.g., clinicians, patients)? If so, will the type of explanation provided by this method be sufficient to ensure comprehension of and control over model predictions, if these are deployed by clinicians? [\textbf{transparency}, \textbf{responsible scientific inference}, \textbf{justice}, \textbf{beneficence}]
\item If the representation of different groups in the training data is not balanced, what measures are in place to ensure models do not become biased towards overrepresented groups or neglect the needs of underrepresented ones?  [\textbf{justice}]
\end{itemize}

\subsection{Evaluation Protocols}
The selection of evaluation protocols should ensure that performance metrics and data used for evaluation truly reflect the model's applicability in real-world mental health care scenarios. In this respect, with regards to principles highlighted above, relevant questions are the following:
\begin{itemize}
\item Which metrics and data are selected to ensure that models are evaluated in a way that provides reliable priors on expected performance in real-world scenarios? What ethical considerations guide the choice of these metrics and data? [\textbf{responsible scientific inference}, \textbf{social responsibility}, \textbf{transparency}]
\item Which evaluation practices are in place to ensure that models' performance on different population subgroups (and related disparities) are thoroughly evaluated? [\textbf{justice}, \textbf{transparency}]
\item Can any additional analysis be performed to assess how the model contributes to a deeper understanding of the relation between language and neuropsychiatric conditions? [\textbf{social responsibility}, \textbf{beneficence}, \textbf{responsible scientific inference}]
\item Which evaluation protocols are in place to identify in which scenarios the model is likely to provide unreliable predictions, so to explicitly discourage specific use cases or deployment on specific populations? [\textbf{beneficence}, \textbf{justice}, \textbf{transparency}, \textbf{responsible scientific inference}]
\item If the model is initialized from pretrained checkpoints, which protocols are in place to evaluated whether any biases are embedded in the initial weights? How is this information used to inform the choice of pretrained model, or devise training or deployment protocols that might mitigate these initial biases? [\textbf{justice}, \textbf{transparency}]
\item Which explainability criteria and techniques are employed to guarantee transparency and understanding of the model's behavior and limitations, ensuring stakeholders can trust and verify its decisions? [\textbf{transparency}, \textbf{responsible scientific inference}, \textbf{beneficence}, \textbf{social responsibility}]
\end{itemize}

\subsection{Model and Data Availability}
With respect to choices concerning sharing of models and data, the following questions emerge from the ethical principles highlighted above:
\begin{itemize}
\item How do we balance the need for transparency and reproducibility with the protection of confidential, sensitive or proprietary information? Which data and model sharing protocols can be used, and how can the data be processed to achieve an optimal trade-off? [\textbf{autonomy}, \textbf{transparency}] 
\item How can models and data be shared so to minimize the likelihood of dual and unintended uses? [\textbf{beneficence}, \textbf{social responsibility}]
\item How can training, evaluation procedures, models and data be shared to ensure full transparency in disclosing methods and results? [\textbf{transparency}] 
\item What measures are in place (e.g., model compression) to allow for the replication of the results and re-use of the model in cases in which the computational burden is substantial? [\textbf{responsible scientific inference}, \textbf{justice}, \textbf{transparency}, \textbf{beneficence}, \textbf{social responsibility}]
\end{itemize}

\subsection{Mode of Deployment}
Finally, while deployment is partly extrinsic to the model development pipeline, the principles discussed above -- especially those of \textbf{beneficence} and \textbf{social responsibility} --  highlight the need to consider broader questions related to applicability \textit{during} model development. In this respect relevant questions are the following:
\begin{itemize}
\item  How can patients' be informed in an accessible and transparent way of how the model produces predictions, and how these are used in the context of a human-centered decision-making process? [\textbf{autonomy}, \textbf{transparency}]
\item How can we envisage that models are embedded in clinical decision-making structures that maximize the good of patients and minimize potential harm? How will the autonomy of patients, clinicians, and care providers be respected in the context of such structures? [\textbf{autonomy}, \textbf{beneficence}, \textbf{social responsibility}]
\item How are explainability techniques embedded in this process, and how are they served to users (e.g., clinicians) to ensure safety and trustworthiness? [\textbf{transparency}, \textbf{beneficence}, \textbf{social responsibility}]
\item How can patients and clinicians be informed about the potential shortcomings of the predictive modeling system? [\textbf{transparency}]
\item How can the cost and likelihood of errors be communicated to stakeholders, especially clinicians? [\textbf{transparency}, \textbf{beneficence}, \textbf{social responsibility}]
\item What strategies are implemented to guarantee that the benefits involved in the deployment of ML models in neuropsychiatric diagnosis are fair and distributed across all relevant population segments, and do not exacerbate existing disparities? [\textbf{justice}]
\end{itemize}

While this list of questions is not meant to provide a comprehensive overview of the ethical landscape of ML research on language-based inference of psychiatric conditions (and while not all questions might be relevant to all research endeavors), we believe that this toolkit can serve as a concrete starting point for ethical analysis and reflection, to be iteratively improved upon in the future. As the formulation of some of these questions remains general to cover a broad variety of use cases, we provide an example of a case study that showcase its utility in a concrete implementation.

\section{Case Study: Linguistic Markers of Autism}

To provide a concrete example of how to apply this toolkit, let us take research on language markers of autism as an example. A quick search on Google Scholar finds tens of dedicated articles \cite{fusaroli2017voice}, an InterSpeech challenge on identifying reliable markers \cite{asgari2013robust}, and initiatives to validate proposed markers \cite{cho-etal-2022-identifying, fusaroli2022toward, hansen_speech-_2023, rybner_vocal_2022}. The interest is clearly there, less clear is whether an in-depth ethical reflection has been performed.

If we consider \textbf{target variables}, we can see that most studies address the presence of autism against controls without neurodevelopmental conditions (e.g. ADHD, specific language impairments, social anxiety), and in a few cases ADOS scores \cite{fusaroli2017voice, fusaroli2022toward}. To justify this (often convenient) choice, most published papers mention the need for prompter and more accurate diagnoses. 

However, such need is far from well-established in the literature. First, no actual evidence has been produced about the often vented benefits of earlier diagnoses \cite{fletcher2024name}. Further, an autism diagnosis is not a clear biological construct, but in important ways a pragmatic tool, which can grant access to services, identity and recognition.

A direct engagement of stakeholders might provide alternative target variables that are better aligned with intended real-world benefits (e.g., supporting clinical practices, and improving outcomes for people with autism; cf. \textbf{beneficence, autonomy, social responsibility}). 
For example, from a clinician's perspective, diagnostic assessment requires the evaluation of multiple possible diagnoses. This involves assessing whether a language delay or other signs of possible neurodevelopmental atypicality are due to an autism or another condition, with different paths for intervention. Further, the high co-morbidity of multiple conditions (e.g. ADHD, autism, social anxiety, \cite{matson2013comorbidity}) suggests that a dichotomous diagnosis is not very useful. Practitioners would rather use scaffolding and facilitation in the assessment of clinical and cognitive traits -- which could be maximally informative of the presence of the one or the other condition, or their co-presence but are time-consuming to perform and to reliably agree on across assessors. For instance, to help disentangle co-morbidities, one could try to infer the presence of anxiety, or lack of attentional focus in the voice, turn-taking and speech content, rather than a diagnostic label. Involving autistic individuals and their communities would surely bring further alternative foci for target variables in focus, and help identify the risks of maleficent dual use for each candidate target (e.g., using models for inference of diagnostic labels for screening in job interviews).

Assuming one or more relevant targets have been identified, the next question is which kind of \textbf{training data} should be assembled. A key concern is the heterogeneity of individuals with autism (\textbf{justice}). Currently more male at birth individuals are diagnosed, compared to female at birth, and more white middle class individuals -- which are also more likely to participate in research -- than individuals from other groups. This makes collecting data on white middle class boys and men much easier than on other groups, which generally leads to unbalanced training and test data, and likely biases in the algorithm. 

Further, given the non-trivial presence of non-native and multilingual speakers in most societies, the current focus of training data on monolingual speakers (which is aimed to ensure a higher homogeneity in the sample and fewer potential confounds in traditional statistical settings) seems problematic. Autism and non-monolingualism are in fact likely to interact \cite{gilhuber2023multiling}. Furthermore, while having to include large and diverse samples might be a practical and economic concern for researchers, these are needed by contemporary NLP approaches. Ignoring the heterogeneity of autism is not an option if we truly want to develop robust and fair models that can be useful in clinical settings. 

The availability of large and heterogeneous samples also feed into considerations regarding \textbf{evaluation protocols} to be planned and implemented. Along the lines of the considerations presented above, it becomes apparent that models should be evaluated for their accuracy across parameters which include monolingual/multilingual status, demographic, and socio-economic characteristics. Importantly, definitions and measures of fairness tend to be problematic if applied independently of the context \cite{corbett2023measure, mcdermott2024closer}. Thus, one need to consult with clinicians and communities to identify differential costs and benefits of misassessment for the different groups, and accordingly stratify evaluations and error analyses.

Further, relevant considerations concern the choice of \textbf{model architectures} in the light of the need for explainability. As the evaluation protocols identify problematic areas to be improved upon, improvement can be better developed if one is able to access confidence of the model and explanations of the decisions made. Explainability should not only to serve the analyst's needs. It should also support clinicians and patients (\textbf{modes of deployment}). A key question, in this respect, is that of how to enable the algorithm to empower clinicians in focusing on the needs of the patient, rather than reduce their agency and centrality in decision-making.

This short case study is an initial example of how questions raised by ELLIPS can be used to develop models that can directly target clinicians' and patients' needs. We hope that researchers will benefit from this initial attempt at charting the ethical landscape of LLM-based research on inference of neuropsychiatric conditions, and utilize our toolkit to improve the alignment between real-world goals of their research and technical decisions made at all stages of model development and deployment.

\section{Conclusion}
While the advent of LLMs has opened up vast opportunities for research on language-based inference of neuropsychiatric conditions, existing research has hitherto failed to deliver impactful and applicable models. As argued in the present paper, this partly arises from a failure to engage with ethical questions related to the alignment of predictive models with their intended applications.

In this paper, we have provided a first attempt at charting the complex ethical landscape of research on language-based inference of neuropsychiatric conditions. We have especially focused on identifying general ethical principles that should guide decisions related to model development and deployment, and translated them into ELLIPS, a toolkit consisting of a set of concrete questions that can guide researchers in their process of aligning their systems with such principles and with their intended real-world application.

With this we hope to encourage practices that can accelerate the emergence of models with positive impact on mental health care, and set an example for other domains within applied machine learning and natural language processing.

\section{Acknowledgements}
The authors would like to thank Katie Link for her work on an early version of the manuscript, and the Interacting Minds Centre for hosting the seminar that initiated this collaboration.

\bibliography{aaai24}

\begin{thebibliography}{56}
\providecommand{\natexlab}[1]{#1}

\bibitem[{Asgari, Bayestehtashk, and Shafran(2013)}]{asgari2013robust}
Asgari, M.; Bayestehtashk, A.; and Shafran, I. 2013.
\newblock Robust and accurate features for detecting and diagnosing autism spectrum disorders.
\newblock In \emph{Interspeech}, volume 2013, 191. NIH Public Access.

\bibitem[{Baird and Schuller(2020)}]{baird2020considerations}
Baird, A.; and Schuller, B. 2020.
\newblock Considerations for a more ethical approach to data in AI: On data representation and infrastructure.
\newblock \emph{Frontiers in big Data}, 3: 25.

\bibitem[{Beauchamp and Childress(1979)}]{beauchamp1978principles}
Beauchamp, T.~L.; and Childress, J.~F. 1979.
\newblock \emph{Principles of biomedical ethics}.
\newblock Oxford University Press, USA.

\bibitem[{Bosma and Granger(2022)}]{bosma2022sharing}
Bosma, C.~M.; and Granger, A.~M. 2022.
\newblock Sharing is caring: Ethical implications of transparent research in psychology.
\newblock \emph{American Psychologist}, 77(4): 565.

\bibitem[{Bueter(2019)}]{bueter2019epistemic}
Bueter, A. 2019.
\newblock Epistemic injustice and psychiatric classification.
\newblock \emph{Philosophy of Science}, 86(5): 1064--1074.

\bibitem[{Cho et~al.(2022)Cho, Fusaroli, Pelella, Tena, Knox, Hauptmann, Covello, Russell, Miller, Hulink, Uzokwe, Walker, Fiumara, Pandey, Chatham, Cieri, Schultz, Liberman, and Parish-morris}]{cho-etal-2022-identifying}
Cho, S.; Fusaroli, R.; Pelella, M.~R.; Tena, K.; Knox, A.; Hauptmann, A.; Covello, M.; Russell, A.; Miller, J.; Hulink, A.; Uzokwe, J.; Walker, K.; Fiumara, J.; Pandey, J.; Chatham, C.; Cieri, C.; Schultz, R.; Liberman, M.; and Parish-morris, J. 2022.
\newblock Identifying stable speech-language markers of autism in children: Preliminary evidence from a longitudinal telephony-based study.
\newblock In Zirikly, A.; Atzil-Slonim, D.; Liakata, M.; Bedrick, S.; Desmet, B.; Ireland, M.; Lee, A.; MacAvaney, S.; Purver, M.; Resnik, R.; and Yates, A., eds., \emph{Proceedings of the Eighth Workshop on Computational Linguistics and Clinical Psychology}, 40--46. Seattle, USA: Association for Computational Linguistics.

\bibitem[{Ciampelli et~al.(2023)Ciampelli, Voppel, de~Boer, Koops, and Sommer}]{ciampelli2023combining}
Ciampelli, S.; Voppel, A.; de~Boer, J.; Koops, S.; and Sommer, I. 2023.
\newblock Combining automatic speech recognition with semantic natural language processing in schizophrenia.
\newblock \emph{Psychiatry Research}, 325: 115252.

\bibitem[{Conway et~al.(2019)Conway, Forbes, Forbush, Fried, Hallquist, Kotov, Mullins-Sweatt, Shackman, Skodol, South, Sunderland, Waszczuk, Zald, Afzali, Bornovalova, Carragher, Docherty, Jonas, Krueger, Patalay, Pincus, Tackett, Reininghaus, Waldman, Wright, Zimmermann, Bach, Bagby, Chmielewski, Cicero, Clark, Dalgleish, DeYoung, Hopwood, Ivanova, Latzman, Patrick, Ruggero, Samuel, Watson, and Eaton}]{conway_hierarchical_2019}
Conway, C.~C.; Forbes, M.~K.; Forbush, K.~T.; Fried, E.~I.; Hallquist, M.~N.; Kotov, R.; Mullins-Sweatt, S.~N.; Shackman, A.~J.; Skodol, A.~E.; South, S.~C.; Sunderland, M.; Waszczuk, M.~A.; Zald, D.~H.; Afzali, M.~H.; Bornovalova, M.~A.; Carragher, N.; Docherty, A.~R.; Jonas, K.~G.; Krueger, R.~F.; Patalay, P.; Pincus, A.~L.; Tackett, J.~L.; Reininghaus, U.; Waldman, I.~D.; Wright, A. G.~C.; Zimmermann, J.; Bach, B.; Bagby, R.~M.; Chmielewski, M.; Cicero, D.~C.; Clark, L.~A.; Dalgleish, T.; DeYoung, C.~G.; Hopwood, C.~J.; Ivanova, M.~Y.; Latzman, R.~D.; Patrick, C.~J.; Ruggero, C.~J.; Samuel, D.~B.; Watson, D.; and Eaton, N.~R. 2019.
\newblock A {Hierarchical} {Taxonomy} of {Psychopathology} {Can} {Transform} {Mental} {Health} {Research}.
\newblock \emph{Perspectives on Psychological Science}, 14(3): 419--436.

\bibitem[{Corbett-Davies et~al.(2023)Corbett-Davies, Gaebler, Nilforoshan, Shroff, and Goel}]{corbett2023measure}
Corbett-Davies, S.; Gaebler, J.~D.; Nilforoshan, H.; Shroff, R.; and Goel, S. 2023.
\newblock The measure and mismeasure of fairness.
\newblock \emph{The Journal of Machine Learning Research}, 24(1): 14730--14846.

\bibitem[{Corona~Hernández et~al.(2023)Corona~Hernández, Corcoran, Achim, de~Boer, Boerma, Brederoo, Cecchi, Ciampelli, Elvevåg, Fusaroli, Giordano, Hauglid, van Hessen, Hinzen, Homan, de~Kloet, Koops, Kuperberg, Maheshwari, Mota, Parola, Rocca, Sommer, Truong, Voppel, van Vugt, Wijnen, and Palaniyappan}]{corona_hernandez_natural_2023}
Corona~Hernández, H.; Corcoran, C.; Achim, A.~M.; de~Boer, J.~N.; Boerma, T.; Brederoo, S.~G.; Cecchi, G.~A.; Ciampelli, S.; Elvevåg, B.; Fusaroli, R.; Giordano, S.; Hauglid, M.; van Hessen, A.; Hinzen, W.; Homan, P.; de~Kloet, S.~F.; Koops, S.; Kuperberg, G.~R.; Maheshwari, K.; Mota, N.~B.; Parola, A.; Rocca, R.; Sommer, I. E.~C.; Truong, K.; Voppel, A.~E.; van Vugt, M.; Wijnen, F.; and Palaniyappan, L. 2023.
\newblock Natural {Language} {Processing} {Markers} for {Psychosis} and {Other} {Psychiatric} {Disorders}: {Emerging} {Themes} and {Research} {Agenda} {From} a {Cross}-{Linguistic} {Workshop}.
\newblock \emph{Schizophrenia Bulletin}, 49(Supplement\_2): S86--S92.

\bibitem[{Devlin et~al.(2019)Devlin, Chang, Lee, and Toutanova}]{devlin_bert_2019}
Devlin, J.; Chang, M.-W.; Lee, K.; and Toutanova, K. 2019.
\newblock {BERT}: {Pre}-training of {Deep} {Bidirectional} {Transformers} for {Language} {Understanding}.
\newblock In \emph{Proceedings of the 2019 {Conference} of the {North} {American} {Chapter} of the {Association} for {Computational} {Linguistics}: {Human} {Language} {Technologies}, {Volume} 1 ({Long} and {Short} {Papers})}, 4171--4186. Minneapolis, Minnesota: Association for Computational Linguistics.

\bibitem[{Douglas(2014)}]{douglas2014moral}
Douglas, H. 2014.
\newblock The moral terrain of science.
\newblock \emph{Erkenntnis}, 79: 961--979.

\bibitem[{Elvev{\aa}g(2023)}]{elvevaag2023reflections}
Elvev{\aa}g, B. 2023.
\newblock Reflections on measuring disordered thoughts as expressed via language.
\newblock \emph{Psychiatry Research}, 322: 115098.

\bibitem[{Fletcher-Watson(2024)}]{fletcher2024name}
Fletcher-Watson, S. 2024.
\newblock What’s in a name? The costs and benefits of a formal autism diagnosis.
\newblock \emph{Autism}, 28(2): 257--262.
\newblock PMID: 37997793.

\bibitem[{Fried and Nesse(2015)}]{fried_depression_2015}
Fried, E.~I.; and Nesse, R.~M. 2015.
\newblock Depression is not a consistent syndrome: an investigation of unique symptom patterns in the {STAR}*{D} study.
\newblock \emph{Journal of affective disorders}, 172: 96--102.

\bibitem[{Fusaroli et~al.(2022)Fusaroli, Grossman, Bilenberg, Cantio, Jepsen, and Weed}]{fusaroli2022toward}
Fusaroli, R.; Grossman, R.; Bilenberg, N.; Cantio, C.; Jepsen, J. R.~M.; and Weed, E. 2022.
\newblock Toward a cumulative science of vocal markers of autism: A cross-linguistic meta-analysis-based investigation of acoustic markers in American and Danish autistic children.
\newblock \emph{Autism Research}, 15(4): 653--664.

\bibitem[{Fusaroli et~al.(2017)Fusaroli, Lambrechts, Bang, Bowler, and Gaigg}]{fusaroli2017voice}
Fusaroli, R.; Lambrechts, A.; Bang, D.; Bowler, D.~M.; and Gaigg, S.~B. 2017.
\newblock Is voice a marker for Autism spectrum disorder? A systematic review and meta-analysis.
\newblock \emph{Autism Research}, 10(3): 384--407.

\bibitem[{Galatzer-Levy and Bryant(2013)}]{galatzer-levy_636120_2013}
Galatzer-Levy, I.~R.; and Bryant, R.~A. 2013.
\newblock 636,120 {Ways} to {Have} {Posttraumatic} {Stress} {Disorder}.
\newblock \emph{Perspectives on Psychological Science}, 8(6): 651--662.
\newblock Publisher: SAGE Publications Inc.

\bibitem[{Gilhuber, Raulston, and Galley(2023)}]{gilhuber2023multiling}
Gilhuber, C.~S.; Raulston, T.~J.; and Galley, K. 2023.
\newblock Language and communication skills in multilingual children on the autism spectrum: A systematic review.
\newblock \emph{Autism}, 27(6): 1516--1531.
\newblock PMID: 36629040.

\bibitem[{Grote and Berens(2020)}]{grote2020ethics}
Grote, T.; and Berens, P. 2020.
\newblock On the ethics of algorithmic decision-making in healthcare.
\newblock \emph{Journal of medical ethics}, 46(3): 205--211.

\bibitem[{Hansen et~al.(2023)Hansen, Rocca, Simonsen, Olsen, Parola, Bliksted, Ladegaard, Bang, Tylén, Weed, Østergaard, and Fusaroli}]{hansen_speech-_2023}
Hansen, L.; Rocca, R.; Simonsen, A.; Olsen, L.; Parola, A.; Bliksted, V.; Ladegaard, N.; Bang, D.; Tylén, K.; Weed, E.; Østergaard, S.~D.; and Fusaroli, R. 2023.
\newblock Speech- and text-based classification of neuropsychiatric conditions in a multidiagnostic setting.
\newblock \emph{Nature Mental Health}, 1(12): 971--981.
\newblock Number: 12 Publisher: Nature Publishing Group.

\bibitem[{Insel(2018)}]{insel_digital_2018}
Insel, T. 2018.
\newblock Digital phenotyping: a global tool for psychiatry.
\newblock \emph{World Psychiatry}, 17(3): 276--277.

\bibitem[{Joshi et~al.(2020)Joshi, Santy, Budhiraja, Bali, and Choudhury}]{joshi2020state}
Joshi, P.; Santy, S.; Budhiraja, A.; Bali, K.; and Choudhury, M. 2020.
\newblock The state and fate of linguistic diversity and inclusion in the NLP world.
\newblock \emph{arXiv preprint arXiv:2004.09095}.

\bibitem[{Kircher et~al.(2018)Kircher, Br{\"o}hl, Meier, and Engelen}]{kircher2018formal}
Kircher, T.; Br{\"o}hl, H.; Meier, F.; and Engelen, J. 2018.
\newblock Formal thought disorders: from phenomenology to neurobiology.
\newblock \emph{The Lancet Psychiatry}, 5(6): 515--526.

\bibitem[{Larsson and Heintz(2020)}]{larsson2020transparency}
Larsson, S.; and Heintz, F. 2020.
\newblock Transparency in artificial intelligence.
\newblock \emph{Internet Policy Review}, 9(2).

\bibitem[{Low, Bentley, and Ghosh(2020)}]{low_automated_2020}
Low, D.~M.; Bentley, K.~H.; and Ghosh, S.~S. 2020.
\newblock Automated assessment of psychiatric disorders using speech: {A} systematic review.
\newblock \emph{Laryngoscope Investigative Otolaryngology}, 5(1): 96--116.
\newblock \_eprint: https://onlinelibrary.wiley.com/doi/pdf/10.1002/lio2.354.

\bibitem[{Matson and Goldin(2013)}]{matson2013comorbidity}
Matson, J.~L.; and Goldin, R.~L. 2013.
\newblock Comorbidity and autism: Trends, topics and future directions.
\newblock \emph{Research in autism spectrum disorders}, 7(10): 1228--1233.

\bibitem[{McDermott et~al.(2024)McDermott, Hansen, Zhang, Angelotti, and Gallifant}]{mcdermott2024closer}
McDermott, M.; Hansen, L.~H.; Zhang, H.; Angelotti, G.; and Gallifant, J. 2024.
\newblock A Closer Look at AUROC and AUPRC under Class Imbalance.
\newblock \emph{arXiv preprint arXiv:2401.06091}.

\bibitem[{Mora-Cantallops et~al.(2021)}]{mora2021traceability}
Mora-Cantallops, M.; et~al. 2021.
\newblock Traceability for trustworthy AI: A review of models and tools.
\newblock \emph{Big Data and Cognitive Computing}, 5(2): 20.

\bibitem[{Nasr et~al.(2023)Nasr, Carlini, Hayase, Jagielski, Cooper, Ippolito, Choquette-Choo, Wallace, Tram{\`e}r, and Lee}]{nasr2023scalable}
Nasr, M.; Carlini, N.; Hayase, J.; Jagielski, M.; Cooper, A.~F.; Ippolito, D.; Choquette-Choo, C.~A.; Wallace, E.; Tram{\`e}r, F.; and Lee, K. 2023.
\newblock Scalable extraction of training data from (production) language models.
\newblock \emph{arXiv preprint arXiv:2311.17035}.

\bibitem[{Nguyen et~al.(2014)Nguyen, Phung, Dao, Venkatesh, and Berk}]{nguyen_affective_2014}
Nguyen, T.; Phung, D.; Dao, B.; Venkatesh, S.; and Berk, M. 2014.
\newblock Affective and {Content} {Analysis} of {Online} {Depression} {Communities}.
\newblock \emph{IEEE Transactions on Affective Computing}, 5(3): 217--226.
\newblock Conference Name: IEEE Transactions on Affective Computing.

\bibitem[{of~the Psychiatric Genomics~Consortium(2019)}]{cross-disorder_group_of_the_psychiatric_genomics_consortium_electronic_address_plee0mghharvardedu_genomic_2019}
of~the Psychiatric Genomics~Consortium, C.-D.~G. 2019.
\newblock Genomic {Relationships}, {Novel} {Loci}, and {Pleiotropic} {Mechanisms} across {Eight} {Psychiatric} {Disorders}.
\newblock \emph{Cell}, 179(7): 1469--1482.e11.

\bibitem[{Panch, Mattie, and Atun(2019)}]{panch2019artificial}
Panch, T.; Mattie, H.; and Atun, R. 2019.
\newblock Artificial intelligence and algorithmic bias: implications for health systems.
\newblock \emph{Journal of global health}, 9(2).

\bibitem[{Parikh, Teeple, and Navathe(2019)}]{parikh2019addressing}
Parikh, R.~B.; Teeple, S.; and Navathe, A.~S. 2019.
\newblock Addressing bias in artificial intelligence in health care.
\newblock \emph{Jama}, 322(24): 2377--2378.

\bibitem[{Park et~al.(2015)Park, Schwartz, Eichstaedt, Kern, Kosinski, Stillwell, Ungar, and Seligman}]{park_automatic_2015}
Park, G.; Schwartz, H.~A.; Eichstaedt, J.~C.; Kern, M.~L.; Kosinski, M.; Stillwell, D.~J.; Ungar, L.~H.; and Seligman, M. E.~P. 2015.
\newblock Automatic personality assessment through social media language.
\newblock \emph{Journal of Personality and Social Psychology}, 108(6): 934--952.
\newblock Place: US Publisher: American Psychological Association.

\bibitem[{Parola et~al.(2023)Parola, Lin, Simonsen, Bliksted, Zhou, Wang, Inoue, Koelkebeck, and Fusaroli}]{parola2023speech}
Parola, A.; Lin, J.~M.; Simonsen, A.; Bliksted, V.; Zhou, Y.; Wang, H.; Inoue, L.; Koelkebeck, K.; and Fusaroli, R. 2023.
\newblock Speech disturbances in schizophrenia: Assessing cross-linguistic generalizability of NLP automated measures of coherence.
\newblock \emph{Schizophrenia Research}, 259: 59--70.

\bibitem[{Peled(2018)}]{peled2018language}
Peled, Y. 2018.
\newblock Language barriers and epistemic injustice in healthcare settings.
\newblock \emph{Bioethics}, 32(6): 360--367.

\bibitem[{Plana-Ripoll et~al.(2019{\natexlab{a}})Plana-Ripoll, Pedersen, Agerbo, Holtz, Erlangsen, Canudas-Romo, Andersen, Charlson, Christensen, Erskine, Ferrari, Iburg, Momen, Mortensen, Nordentoft, Santomauro, Scott, Whiteford, Weye, McGrath, and Laursen}]{plana-ripoll_comprehensive_2019}
Plana-Ripoll, O.; Pedersen, C.~B.; Agerbo, E.; Holtz, Y.; Erlangsen, A.; Canudas-Romo, V.; Andersen, P.~K.; Charlson, F.~J.; Christensen, M.~K.; Erskine, H.~E.; Ferrari, A.~J.; Iburg, K.~M.; Momen, N.; Mortensen, P.~B.; Nordentoft, M.; Santomauro, D.~F.; Scott, J.~G.; Whiteford, H.~A.; Weye, N.; McGrath, J.~J.; and Laursen, T.~M. 2019{\natexlab{a}}.
\newblock A comprehensive analysis of mortality-related health metrics associated with mental disorders: a nationwide, register-based cohort study.
\newblock \emph{The Lancet}, 394(10211): 1827--1835.

\bibitem[{Plana-Ripoll et~al.(2019{\natexlab{b}})Plana-Ripoll, Pedersen, Holtz, Benros, Dalsgaard, de~Jonge, Fan, Degenhardt, Ganna, Greve, Gunn, Iburg, Kessing, Lee, Lim, Mors, Nordentoft, Prior, Roest, Saha, Schork, Scott, Scott, Stedman, Sørensen, Werge, Whiteford, Laursen, Agerbo, Kessler, Mortensen, and McGrath}]{plana-ripoll_exploring_2019}
Plana-Ripoll, O.; Pedersen, C.~B.; Holtz, Y.; Benros, M.~E.; Dalsgaard, S.; de~Jonge, P.; Fan, C.~C.; Degenhardt, L.; Ganna, A.; Greve, A.~N.; Gunn, J.; Iburg, K.~M.; Kessing, L.~V.; Lee, B.~K.; Lim, C. C.~W.; Mors, O.; Nordentoft, M.; Prior, A.; Roest, A.~M.; Saha, S.; Schork, A.; Scott, J.~G.; Scott, K.~M.; Stedman, T.; Sørensen, H.~J.; Werge, T.; Whiteford, H.~A.; Laursen, T.~M.; Agerbo, E.; Kessler, R.~C.; Mortensen, P.~B.; and McGrath, J.~J. 2019{\natexlab{b}}.
\newblock Exploring {Comorbidity} {Within} {Mental} {Disorders} {Among} a {Danish} {National} {Population}.
\newblock \emph{JAMA Psychiatry}, 76(3): 259--270.

\bibitem[{Rocca and Yarkoni(2021)}]{rocca_putting_2021}
Rocca, R.; and Yarkoni, T. 2021.
\newblock Putting {Psychology} to the {Test}: {Rethinking} {Model} {Evaluation} {Through} {Benchmarking} and {Prediction}.
\newblock \emph{Advances in Methods and Practices in Psychological Science}, 4(3): 25152459211026864.
\newblock Publisher: SAGE Publications Inc.

\bibitem[{Rocca and Yarkoni(2022)}]{rocca_language_2022}
Rocca, R.; and Yarkoni, T. 2022.
\newblock Language as a fingerprint: {Self}-supervised learning of user encodings using transformers.
\newblock In \emph{Findings of the {Association} for {Computational} {Linguistics}: {EMNLP} 2022}, 1701--1714.

\bibitem[{Roefs et~al.(2022)Roefs, Fried, Kindt, Martijn, Elzinga, Evers, Wiers, Borsboom, and Jansen}]{roefs_new_2022}
Roefs, A.; Fried, E.~I.; Kindt, M.; Martijn, C.; Elzinga, B.; Evers, A. W.~M.; Wiers, R.~W.; Borsboom, D.; and Jansen, A. 2022.
\newblock A new science of mental disorders: {Using} personalised, transdiagnostic, dynamical systems to understand, model, diagnose and treat psychopathology.
\newblock \emph{Behaviour Research and Therapy}, 153: 104096.

\bibitem[{Ruder et~al.(2019)Ruder, Peters, Swayamdipta, and Wolf}]{ruder_transfer_2019}
Ruder, S.; Peters, M.~E.; Swayamdipta, S.; and Wolf, T. 2019.
\newblock Transfer {Learning} in {Natural} {Language} {Processing}.
\newblock In \emph{Proceedings of the 2019 {Conference} of the {North} {American} {Chapter} of the {Association} for {Computational} {Linguistics}: {Tutorials}}, 15--18. Minneapolis, Minnesota: Association for Computational Linguistics.

\bibitem[{Rybner et~al.(2022)Rybner, Jessen, Mortensen, Larsen, Grossman, Bilenberg, Cantio, Jepsen, Weed, and Simonsen}]{rybner_vocal_2022}
Rybner, A.; Jessen, E.~T.; Mortensen, M.~D.; Larsen, S.~N.; Grossman, R.; Bilenberg, N.; Cantio, C.; Jepsen, J. R.~M.; Weed, E.; and Simonsen, A. 2022.
\newblock Vocal markers of autism: {Assessing} the generalizability of machine learning models.
\newblock \emph{Autism Research}, 15(6): 1018--1030.
\newblock Publisher: Wiley Online Library.

\bibitem[{Saxena(2018)}]{saxena_excess_2018}
Saxena, S. 2018.
\newblock Excess mortality among people with mental disorders: a public health priority.
\newblock \emph{The Lancet Public Health}, 3(6): e264--e265.

\bibitem[{Shmueli(2010)}]{shmueli_explain_2010}
Shmueli, G. 2010.
\newblock To {Explain} or to {Predict}?
\newblock \emph{Statistical Science}, 25(3): 289--310.
\newblock Publisher: Institute of Mathematical Statistics.

\bibitem[{Sim et~al.(2006)Sim, Chan, G{\"u}lmezoglu, Evans, and Pang}]{sim2006clinical}
Sim, I.; Chan, A.-W.; G{\"u}lmezoglu, A.~M.; Evans, T.; and Pang, T. 2006.
\newblock Clinical trial registration: transparency is the watchword.
\newblock \emph{The Lancet}, 367(9523): 1631--1633.

\bibitem[{Thornicroft(2013)}]{thornicroft_premature_2013}
Thornicroft, G. 2013.
\newblock Premature death among people with mental illness.
\newblock \emph{BMJ}, 346: f2969.
\newblock Publisher: British Medical Journal Publishing Group Section: Editorial.

\bibitem[{Tobia, Nielsen, and Stremitzer(2021)}]{tobia2021does}
Tobia, K.; Nielsen, A.; and Stremitzer, A. 2021.
\newblock When does physician use of AI increase liability?
\newblock \emph{Journal of Nuclear Medicine}, 62(1): 17--21.

\bibitem[{{United Nations Development Programme}(2022)}]{united_nations_development_programme_human_2022}
{United Nations Development Programme}. 2022.
\newblock Human {Development} {Report} 2021-22: {Uncertain} {Times}, {Unsettled} {Lives}: {Shaping} our {Future} in a {Transforming} {World}.
\newblock Technical report.

\bibitem[{Vasudevan et~al.(2022)Vasudevan, Saha, Tarver, and Patel}]{vasudevan_digital_2022}
Vasudevan, S.; Saha, A.; Tarver, M.~E.; and Patel, B. 2022.
\newblock Digital biomarkers: {Convergence} of digital health technologies and biomarkers.
\newblock \emph{npj Digital Medicine}, 5(1): 1--3.
\newblock Number: 1 Publisher: Nature Publishing Group.

\bibitem[{Vaswani et~al.(2017)Vaswani, Shazeer, Parmar, Uszkoreit, Jones, Gomez, Kaiser, and Polosukhin}]{vaswani_attention_2017}
Vaswani, A.; Shazeer, N.; Parmar, N.; Uszkoreit, J.; Jones, L.; Gomez, A.~N.; Kaiser, L.; and Polosukhin, I. 2017.
\newblock Attention is {All} you {Need}.
\newblock In Guyon, I.; Luxburg, U.~V.; Bengio, S.; Wallach, H.; Fergus, R.; Vishwanathan, S.; and Garnett, R., eds., \emph{Advances in {Neural} {Information} {Processing} {Systems}}, volume~30. Curran Associates, Inc.

\bibitem[{Williamson et~al.(2016)Williamson, Godoy, Cha, Schwarzentruber, Khorrami, Gwon, Kung, Dagli, and Quatieri}]{williamson_detecting_2016}
Williamson, J.~R.; Godoy, E.; Cha, M.; Schwarzentruber, A.; Khorrami, P.; Gwon, Y.; Kung, H.-T.; Dagli, C.; and Quatieri, T.~F. 2016.
\newblock Detecting {Depression} using {Vocal}, {Facial} and {Semantic} {Communication} {Cues}.
\newblock In \emph{Proceedings of the 6th {International} {Workshop} on {Audio}/{Visual} {Emotion} {Challenge}}, {AVEC} '16, 11--18. New York, NY, USA: Association for Computing Machinery.
\newblock ISBN 978-1-4503-4516-3.

\bibitem[{{World Health Organization}(2022)}]{world_health_organization_world_nodate}
{World Health Organization}. 2022.
\newblock World mental health report: {Transforming} mental health for all.
\newblock Technical report.

\bibitem[{Yarkoni(2010)}]{yarkoni_personality_2010}
Yarkoni, T. 2010.
\newblock Personality in 100,000 {Words}: {A} large-scale analysis of personality and word use among bloggers.
\newblock \emph{Journal of Research in Personality}, 44(3): 363--373.

\bibitem[{Yarkoni and Westfall(2017)}]{yarkoni_choosing_2017}
Yarkoni, T.; and Westfall, J. 2017.
\newblock Choosing {Prediction} {Over} {Explanation} in {Psychology}: {Lessons} {From} {Machine} {Learning}.
\newblock \emph{Perspectives on Psychological Science: A Journal of the Association for Psychological Science}, 12(6): 1100--1122.

\end{thebibliography}
\end{document}